\newcommand{\linarite} {PbCuSO$_4$(OH)$_2$\xspace}

\documentclass[aps,pra,reprint,showpacs,longbibliography,superscriptaddress, citeonline]{revtex4-1}
\usepackage{amsmath,amssymb,bm}
\usepackage{graphicx}
\usepackage{xspace}
\usepackage{latexsym}
\usepackage{color}
\usepackage{hyperref}
\usepackage{placeins}

\begin{document}
\title{Magnetic phase diagram
of the strongly frustrated quantum spin chain system
PbCuSO$_4$(OH)$_2$ in tilted magnetic fields.}

\author{Y.~Feng}
    \affiliation{Laboratory for Solid State Physics, ETH Z\"{u}rich, 8093 Z\"{u}rich, Switzerland}
    \homepage{http://www.neutron.ethz.ch/}

\author{K.~Yu.~Povarov}
    \email{povarovk@phys.ethz.ch}
    \affiliation{Laboratory for Solid State Physics, ETH Z\"{u}rich, 8093 Z\"{u}rich, Switzerland}

\author{A.~Zheludev}
    \affiliation{Laboratory for Solid State Physics, ETH Z\"{u}rich, 8093 Z\"{u}rich, Switzerland}

\begin{abstract}
We report the $\mathbf{H}-T$ phase diagram of $S=1/2$ strongly
frustrated anisotropic spin chain material linarite PbCuSO$_4$(OH)$_2$ in
tilted magnetic fields up to 10~T and temperatures down to 0.2~K. By
means of torque magnetometry we investigate the phase diagram
evolution as the magnetic field undergoes rotation in
$\mathbf{ba}^{\ast}$ and $\mathbf{bc}$ planes. The key finding is
the robustness of the high field spin density wave-like phase, which
may persist even as the external field goes orthogonal to the chain
direction $\mathbf{b}$. In contrast, the intermediate collinear
antiferromagnetic phase collapses at moderate deflection angles with
respect to $\mathbf{b}$ axis.
\end{abstract}

\date{\today}
\maketitle

\section{Introduction}

Frustrated quantum magnets host extreme quantum fluctuations that
enable a variety of exotic novel phases of spin
matter~\cite{ZhitomirskyHonecker_PRL_2000_FrustratedInfield,ShannonMomoi_PRL_2006_J1J2squarecircle,Sudan_PRB_2009_ChainMultipoles,BalentsStarykh_PRL_2016_QuantumLifshitz}.
Much attention has been given to even the simplest of models, namely
the Heisenberg $S=1/2$ spin chain with ferromagnetic $J_1$ and
next-nearest neighbor antiferromagnetic $J_2$
interactions~\cite{Sudan_PRB_2009_ChainMultipoles,ZhitomirskyTsunetsugu_EPL_2010_nematic,SatoHikihara_PRL_2013_J1J2aniso}.
While quantum fluctuations destabilize the semiclassical spin spiral
order, substantial ferromagnetic interactions favor the formation of
magnon bound states. In applied magnetic fields the bounds states
may condense before single magnons do. The result is the so-called
bond-nematic phase with no dipolar magnetic order, yet spontaneously
broken spin rotational
symmetry~\cite{AndreevGrishchuk_JETP_1984_SpinNematics,ZhitomirskyHonecker_PRL_2000_FrustratedInfield}.
Other unusual quantum phases, such as complicated spiral structures
or spin density waves  (SDW) have also been
predicted~\cite{Sudan_PRB_2009_ChainMultipoles,SatoHikihara_PRL_2013_J1J2aniso,NishimotoDrechsler_PRB_2015_perturbedJ1J2}.

\begin{figure}
  \includegraphics[width=0.5\textwidth]{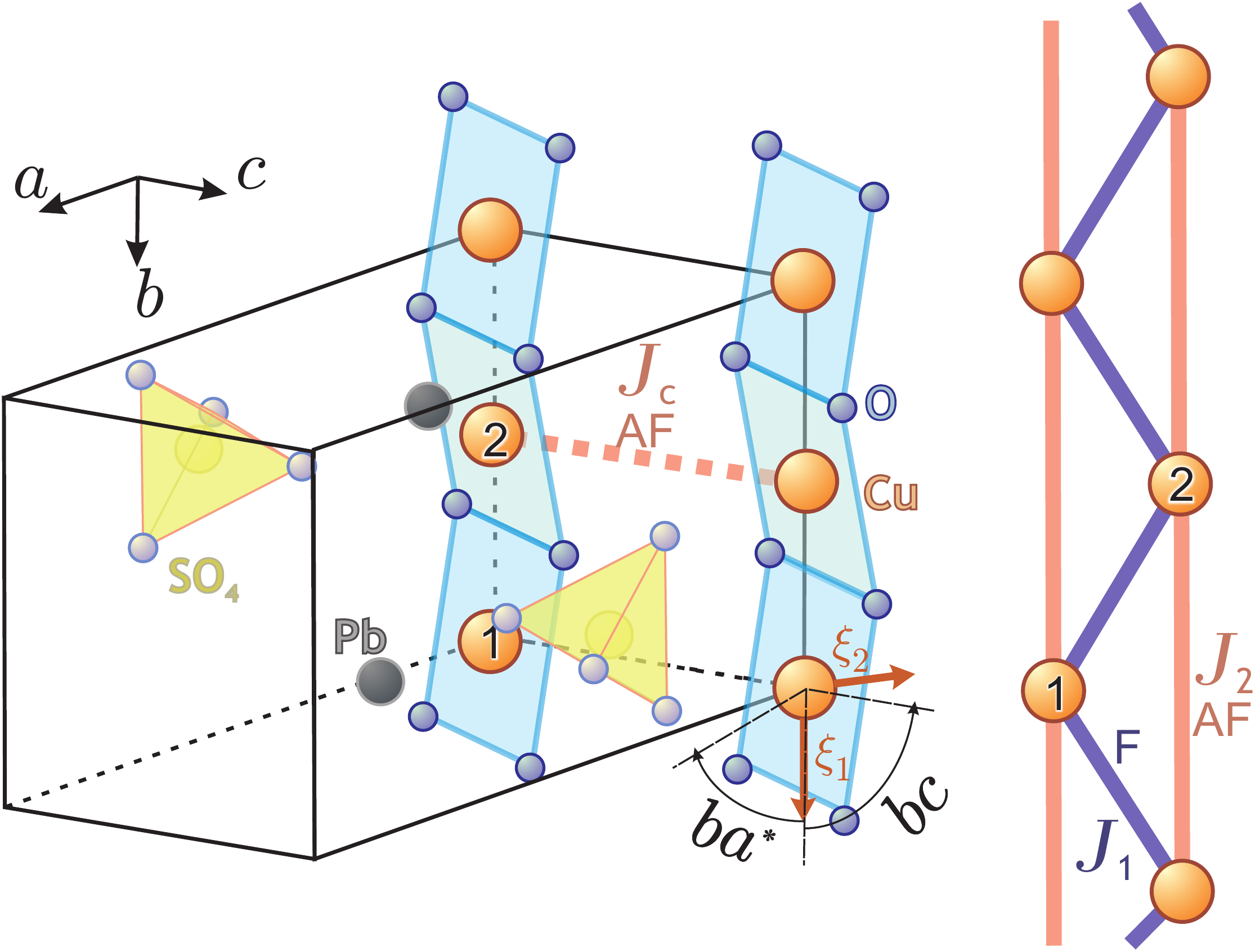}\\
  \caption{The crystal structure of linarite \linarite. The main
  exchange interactions (competing ferromagnetic $J_1$ and antiferromagnetic $J_2$, and interchain $J_{c}$) are indicated.
  The anisotropy axes $\vec{\xi}_{1}\parallel \mathbf{b}$ and $\vec{\xi}_{2}$ in the $\mathbf{ac}$ plane are also shown together with the range of magnetic field
  directions, studied in the present work.}\label{FIG:structure}
\end{figure}

One of the most intriguing potential experimental realizations of
this model
~\cite{Baran_PhysStatSol_2006_LinariteInitiation,Willenberg_PRL_2012_LinariteFrustrated}
is the natural mineral linarite \linarite\ (see
Fig.~\ref{FIG:structure}). It combines pronounced frustration with
very convenient energy scales: in the exchange interactions between
Cu$^{2+}$ $S=1/2$ ions in linarite are $J_{1}\simeq-14.5$ and
$J_{2}\simeq3.93$~meV resulting in a saturation field below $10$~T.
The thermodynamic properties are rather exotic: for the field
applied along the chain direction one finds up to 5 distinct
magnetic phases below
$T_{N}\simeq2.7$~K~\cite{Willenberg_PRL_2016_LinariteSDWs}. Among
them there is especially peculiar high field phase, which was
identified as the longitudinal SDW. The latter was argued to be a
possible precursor to the magnon pair condensate, or possibly even
the phase separation between such a condensate and a conventional
dipolar
order~\cite{Sudan_PRB_2009_ChainMultipoles,Willenberg_PRL_2016_LinariteSDWs}.

Any discussion of linarite in the context of purely isotropic
$J_{1}-J_{2}$ chain
model~\cite{Willenberg_PRL_2016_LinariteSDWs,Rule_PRB_2017_LinariteCNCS}
is incomplete. Magnetic anisotropy certainly plays a role in this
material, as evidenced by the dramatic difference in the phase
diagrams measured for field applied parallel and perpendicular to
the chain
axis~\cite{Schapers_PRB_2013_LinariteBulk,PovarovFeng_PRB_2016_LinariteMF}.
Anisotropy effects were recently addressed in an experimental and
theoretical study~\cite{CemalEnderle_PRL_2018_Linarite}. It was
shown that the magnetically ordered structures can be understood in
terms of mean field model with orthorhombic (biaxial) anisotropy
included. The theoretical description also accounted for a
significant mismatch between the magnetic anisotropy and crystal
lattice directions. The proposed easy and middle axes of the
anisotropy are indicated as $\vec{\xi}_{1}$ and $\vec{\xi}_{2}$
vectors in Fig.~\ref{FIG:structure}. Unfortunately, the available
experimental data~\cite{CemalEnderle_PRL_2018_Linarite} are either
restricted to relatively high temperatures or specific directions of
the magnetic field. A complete orientational low-temperature
magnetic phase diagram of linarite is still lacking.

In the present study we use low-temperature torque magnetometry to map
out this phase diagram for arbitrary magnetic field
directions in $\mathbf{ba}^{\ast}$ and $\mathbf{bc}$ planes.
This allows us to trace the
evolution of each of the magnetic phases as the field is rotated away from the easy
axis direction. Special attention is paid to the high field phase
which we find to be very robust, in contrast with the fragile
intermediate field Ne\'{e}l phase.

\section{Experimental}

\begin{figure}
  \includegraphics[width=0.5\textwidth]{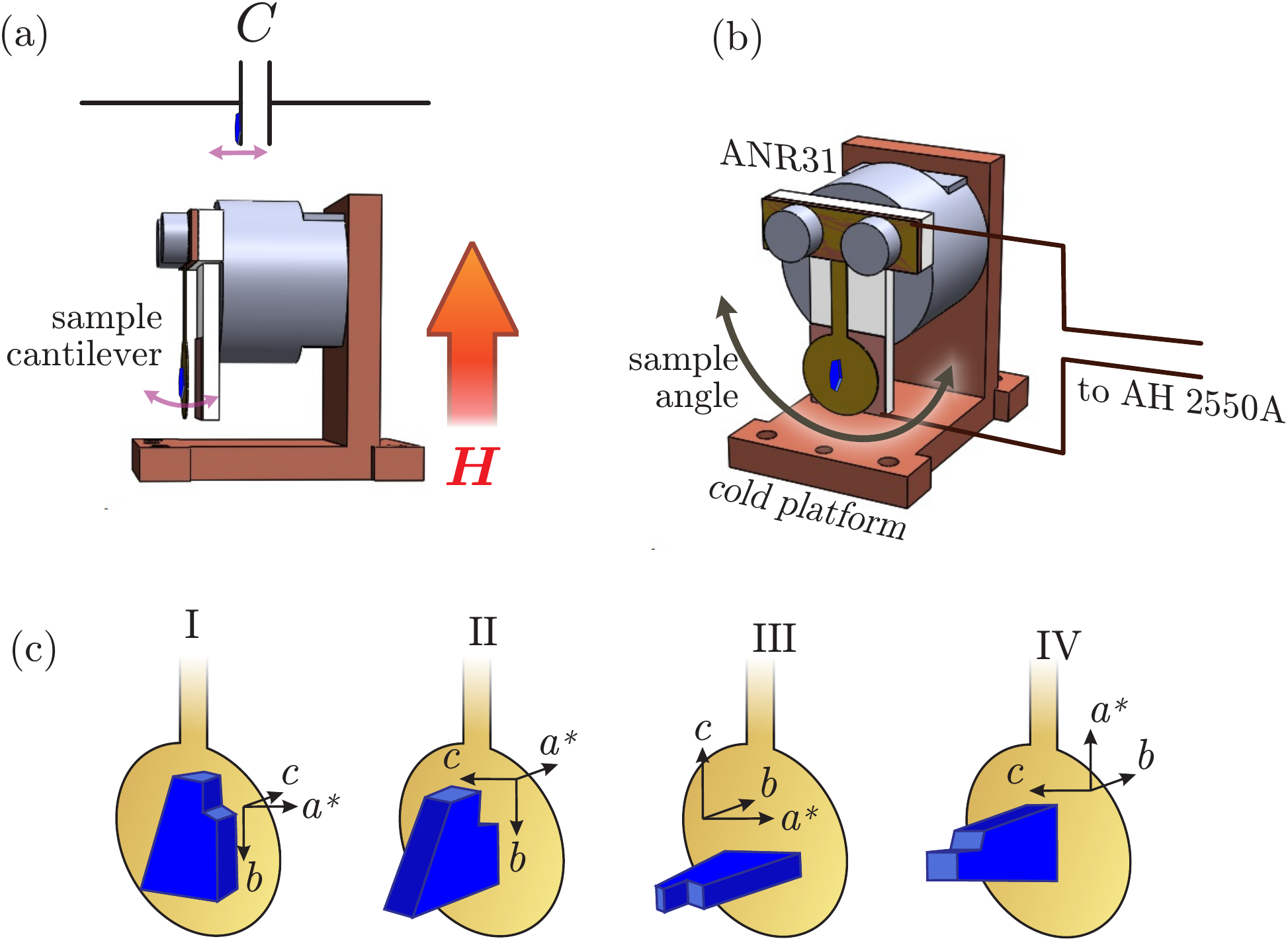}\\
  \caption{Schematics of the custom probe used in the measurements. (a) Principle of the torque measurement:
  probe senses the force, acting on the sample in the vertical magnetic field $\mathbf{H}$,
  by change of the capacitance between the brass cantilever and the
  base copper
  pad
  due to the deflection of the former. A Stycast 1226 panel (shown in white) provides
  an electric insulation between the effective capacitor parts.
  (b) The overview of the probe: the capacitance $C$ of the torque probe is measured by the capacitance bridge AH 2550A, while the angle
  between the field and cantilever principal axis (and hence the sample) can be tuned by the ANR31 rotator (gray). The complete assembly
  is fixed on a copper beryllium rack and
  mounted on the DR cold platform. (c) Four different measurement geometries used in the present study.}\label{FIG:scheme}
\end{figure}

The challenge is to map out the sub-K magnetic phase
diagram of a strongly anisotropic system, featuring many transitions
that substantially affect the magnetization $\mathbf{M}$. This makes
torque magnetometry a very advantageous probe. In our particular
realization the sample is attached to the free end of a cantilever,
$\mathbf{L}$ being the vector from its fixed point to the sample.
Then the torque acting on the cantilever free end consists of two
terms:

\begin{equation}\label{EQ:torquedef}
\mathcal{T}=\mathcal{T}_{\perp}+\mathcal{T}_{\parallel}=[\mathbf{M}\times
\mathbf{H}]+[\mathbf{L}\times(\mathbf{M}\cdot\nabla)\mathbf{H}].
\end{equation}

The $\mathcal{T}_{\perp}$ term depends only on the magnetization
component that is transverse to the magnetic field $\mathbf{H}$. The
other term $\mathcal{T}_{\parallel}$ is mostly sensitive to the
component along the field. Therefore, the method probes the
changes in both longitudinal and transverse components of the
uniform magnetization, and this sensitivity progressively increases
with the external field magnitude. On the down
side, the the field gradient dependence in $\mathcal{T}_{\parallel}$
(which would vary depending on the magnet used or the precise sample
position) makes the data difficult to
interpret quantitatively. As we will show below, for the purposes of this
study this is not a concern, as the transition-related features are
conspicuously pronounced in the data, and a simple qualitative interpretation
is sufficient to reconstruct the phase boundaries.

A schematic of a custom torquemeter probe used in this work is shown in
Figs.~\ref{FIG:scheme}(a,b). The sample is attached to the pad of
the cantilever made of of $0.1$~mm thick brass foil. We measure the
cantilever deflection (i.e. the torque force component normal to the
pad) by observing a change in the electric capacitance $C$ between
the pad and the fixed copper plate. The typical capacitance of the
probe is about $0.5$~pF, and the typical deflection-induced change
is within $1$\% of this value. The capacitance $C$ is measured
directly with the help of Andeen-Hagerling 2550A capacitance bridge.
The probe is in turn mounted onto the Attocube ANR31 rotator,
providing the ability to adjust the angle between the sample and the
magnetic field. The measurement unit (Fig.~\ref{FIG:scheme}b) is
attached to the cold platform of the Quantum Design Dilution
Refrigerator option (DR), that is used in a Quantum Design Physical
Property Measurement System (PPMS) equipped with a 14~T
superconducting magnet. A similar PPMS system with a 9~T magnet was
also used in some of the measurements.

For the study we have used a small $m\simeq 20$~mg natural linarite
single crystal (originating from Grand Reef Mine, Arizona, USA).
This crystal belongs to the same batch as the samples from the
previous study~\cite{PovarovFeng_PRB_2016_LinariteMF}. Although some
mechanically induced shape irregularities, two good facets given by
$\mathbf{bc}$ and $\mathbf{a}^{\ast}\mathbf{c}$ lattice vectors are
present. The linear dimensions of the crystal are approximately
$3\times2\times1$~mm along $\mathbf{b}$, $\mathbf{a}^{\ast}$,
$\mathbf{c}$. The crystal was placed onto the cantilever pad in four
different configurations shown in Fig.~\ref{FIG:scheme}(c). The
adjustment of the rotator position was always done at the room
temperature, as the rotator calibration is temperature-dependent.
Initial positioning of the crystal on the cantilever pad is the
biggest source of experimental uncertainty in the magnetic field
angle. We estimate the offset that may occur during the initial
positioning as not exceeding $\pm3^{\circ}$. This offset is constant
within the series of measurements in a given configuration. The
error resulting from readjusting the rotator angle is negligible in
comparison.

Capacitance $C(H)$ measurements were done at a set of fixed
temperatures (0.2~K lowest) with the magnetic field being swept at
20~Oe/sec.

The intrinsic demagnetizing fields of linarite do not exceed 0.1~T,
and are thus comparable to the typical width of the features that
will be discussed below~\footnote{The saturated magnetization per
mole for $S=1/2$ is below $6000$~cm$^3$G/mol. As the molar volume of
\linarite\ is nearly $76$ cm$^3$/mol, the resulting correction to
the external field is somewhat below $80N$~G, where $N$ is the
geometric coefficient, not exceeding $4\pi$ (the infinite plate
case). Thus, largest possible demagnetization correction is $\Delta
H\simeq0.1$~T.  As the sample is rather 3D, we expect that the
actual coefficients for most of the directions would be closer to
the spherical case $4\pi/3$, reducing the correction even further.}.

\section{Results and discussion}

\subsection{Torque curve and the phase transitions}
\label{SEC:TheFits}

\begin{figure}
  \includegraphics[width=0.5\textwidth]{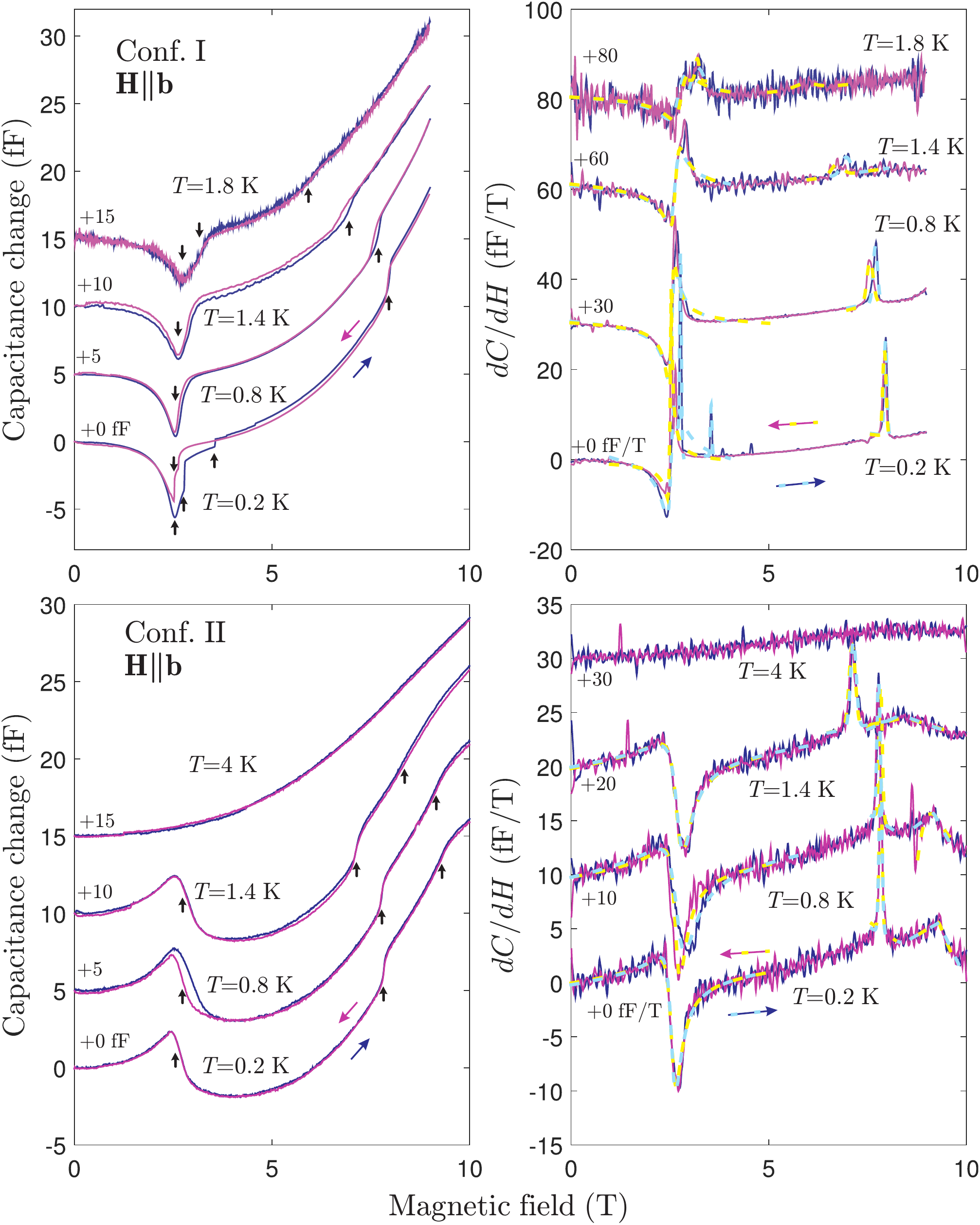}\\
  \caption{Selected $\Delta C(H)$ torque curves and their derivatives for $\mathbf{H}\parallel\mathbf{b}$ in configurations I
  (upper panels) and II (lower panels). The temperatures and offsets
  are indicated in the plots.
  Solid lines are the data; dashed lines are the fits given by Eqs. (\ref{EQ:difflorenzfit},\ref{EQ:gaussfit},\ref{EQ:blurcross}). Vertical arrows mark the
  obtained transition fields. }\label{FIG:Bcurves}
\end{figure}

The raw $\Delta C(H)=C(H)-C(0)$ curves have rather complicated shape
greatly varying depending on the magnetic field direction. The most
structured curves occur at $\mathbf{H}\parallel\mathbf{b}$, the
chain direction. The left panels of Fig.~\ref{FIG:Bcurves} shows the
data, recorded in two different configurations featuring the same
field orientation $\mathbf{H}\parallel \mathbf{b}$. Right panels are
the corresponding $dC/dH$  derivatives. Despite that the curves from
configuration I and II appear very different at the first glance,
they show a number of robust features. These allow us to reproduce
the well-known phase boundaries for $\bf{H}\|\bf{b}$
~\cite{Schapers_PRB_2013_LinariteBulk,Schapers_PRB_2014_LinariteNMR,Willenberg_PRL_2016_LinariteSDWs,PovarovFeng_PRB_2016_LinariteMF,CemalEnderle_PRL_2018_Linarite}.

\begin{figure}
  \includegraphics[width=0.4\textwidth]{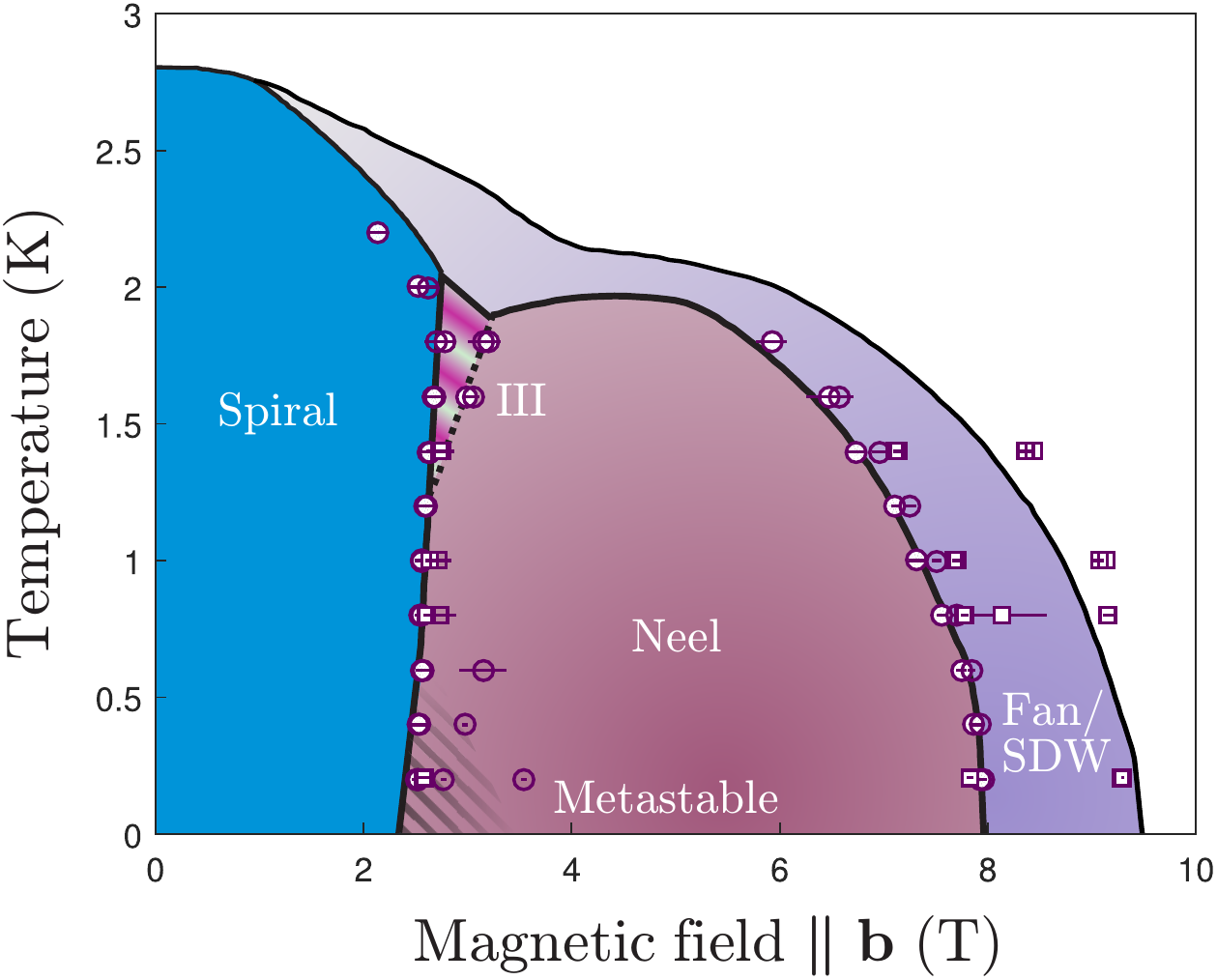}\\
  \caption{Magnetic $\mathbf{H}\parallel \mathbf{b}$ phase diagram of linarite, measured in configurations I (circles) and II (squares).
  The filled and hollow points correspond to down and up field sweeps. The solid lines correspond to the
  previously published phase diagram~\cite{Willenberg_PRL_2016_LinariteSDWs,PovarovFeng_PRB_2016_LinariteMF}. Phases are labeled according to~\cite{CemalEnderle_PRL_2018_Linarite}.}\label{FIG:Bdiag}
\end{figure}

First of all, there is a low field peak-like anomaly (dip or peak
around $H\simeq 3$~T), corresponding to the transition between the
spin spiral and commensurate structure~\footnote{The classification
of the magnetic phases is given according to the most recent
publication~\cite{CemalEnderle_PRL_2018_Linarite}.}. This feature is
rather asymmetric; however, its derivative can be conveniently
described by the distorted Lorentzian function:

\begin{equation}\label{EQ:difflorenzfit}
    \frac{dC}{dH}=aH+b+\frac{I_{0}\sigma^2}{\sigma^{2}+(H-H_{0})^2}\left(1-\frac{\alpha}{\sigma}(H-H_{0}) \right)
\end{equation}

Most of the parameters in the above formula are purely empirical:
the linear background coefficients $a$ and $b$, the anomaly
``amplitude'' $I_{0}$ and the asymmetry coefficient $\alpha$.
Physically meaningful parameters are the peak center $H_{0}$ that is
the transition field and width $\sigma$ that is considered as twice
the experimental uncertainty.

Second, at low temperatures the broad feature may be superimposed
with  abrupt discontinuous jumps, as is the case for $T=0.2$~K curve
in configuration I around $H\simeq 3.5$~T(Fig.~\ref{FIG:Bcurves}).
It is important to note that these jumps \emph{always} have
extremely hysteretic character and are mostly present in the sweeps
with increasing magnetic field. A convenient way of fitting the
jump-like features is to approximate the peak-like derivative with a
Gaussian function, superimposed with linear background:

\begin{equation}\label{EQ:gaussfit}
    \frac{dC}{dH}=aH+b+\frac{I_{0}}{\sigma\sqrt{2\pi}}\text{exp}\left(\frac{-(H-H_{0})^2}{2\sigma^{2}}\right).
\end{equation}

 Again, $a$ and $b$ describe the linear background and $I_{0}$ is the
Gaussian amplitude. Transition field and experimental error are
given by $H_{0}$ and $0.5\sigma$ correspondingly.

The third type of features are the ``smoothed'' jumps, which mark the
lower boundary of the most interesting high field phase. Again, the
derivative of these features is well described by a biased Gaussian
function~(\ref{EQ:gaussfit}).

Finally, the saturation field manifests itself as an apparent kink
in the $\Delta C(H)$ curve. Again, a convenient way to pinpoint the
transition field is an empirical approximation of the derivative
with some peak-like function. Biased Gaussian~(\ref{EQ:gaussfit})
may serve as a good candidate, however we find that in many cases
the ``smoothed angle'' describes the cusp in the derivative more
accurately. It is defined as follows:

\begin{eqnarray}\label{EQ:blurcross}
y(x)&=&a_{1}x+b_{1}, x\leq H_{0},\\ \nonumber
y(x)&=&a_{2}(x-H_{0})+a_{1}H_{0}+b_{1}, x> H_{0},\\
\frac{dC}{dH}&=&\int\limits_{-\infty}^{+\infty}
y(x)\frac{1}{\sigma\sqrt{2\pi}}\text{exp}\left(\frac{-(H-x)^2}{2\sigma^{2}}\right)dx.\nonumber
\end{eqnarray}

The above definition simply describes two straight lines forming a sharp
angle at the anomaly position $H_{0}$, and then convoluted with the
Gaussian of width $\sigma$. As before, this width is a good estimate
for the experimental uncertainty.

In both configurations all features show some temperature
dependence. At $T>T_{N}$ (e.g. $4$~K curve in
Fig.~\ref{FIG:Bcurves}) the $\Delta C(H)$ data become absolutely
featureless, confirming the magnetic order origin of the anomalies
at lower temperatures. Importantly, the highest-field anomaly is
very sensitive to the temperature and becomes almost unobservable
above 1~K. This is a general property of the enigmatic ``Fan/SDW''
phase: it has very weak thermodynamic manifestations at finite $T$
and therefore becomes hardly distinguishable from a fully polarized
state. Empirically this sets $1.4$~K as the threshold temperature at
which this phase of main interest can be resolved.

The result of treating the $\mathbf{H}\parallel \mathbf{b}$ data is
summarized in Fig.~\ref{FIG:Bdiag}. We certainly can reproduce the
entire known phase diagram. The
agreement between the data measured in two different geometries
(configurations I and II) is an additional self-consistency check for our
experimental approach.

\subsection{Evolution in tilted magnetic field}

As the magnetic field gets deflected from the $\mathbf{b}$ axis
towards the $\mathbf{c}$ direction, the torque $\Delta C(H)$ curves
undergo substantial changes. The most obvious but least informative trend is the deformation
of overall shape of the curves.
It largely depends on the multiple geometrical factors in
Eq.~(\ref{EQ:torquedef}) that are at least partially beyond the
experimental control. The really valuable information is contained
in the changing $\Delta C(H)$ anomalies.

\begin{figure}
  \includegraphics[width=0.5\textwidth]{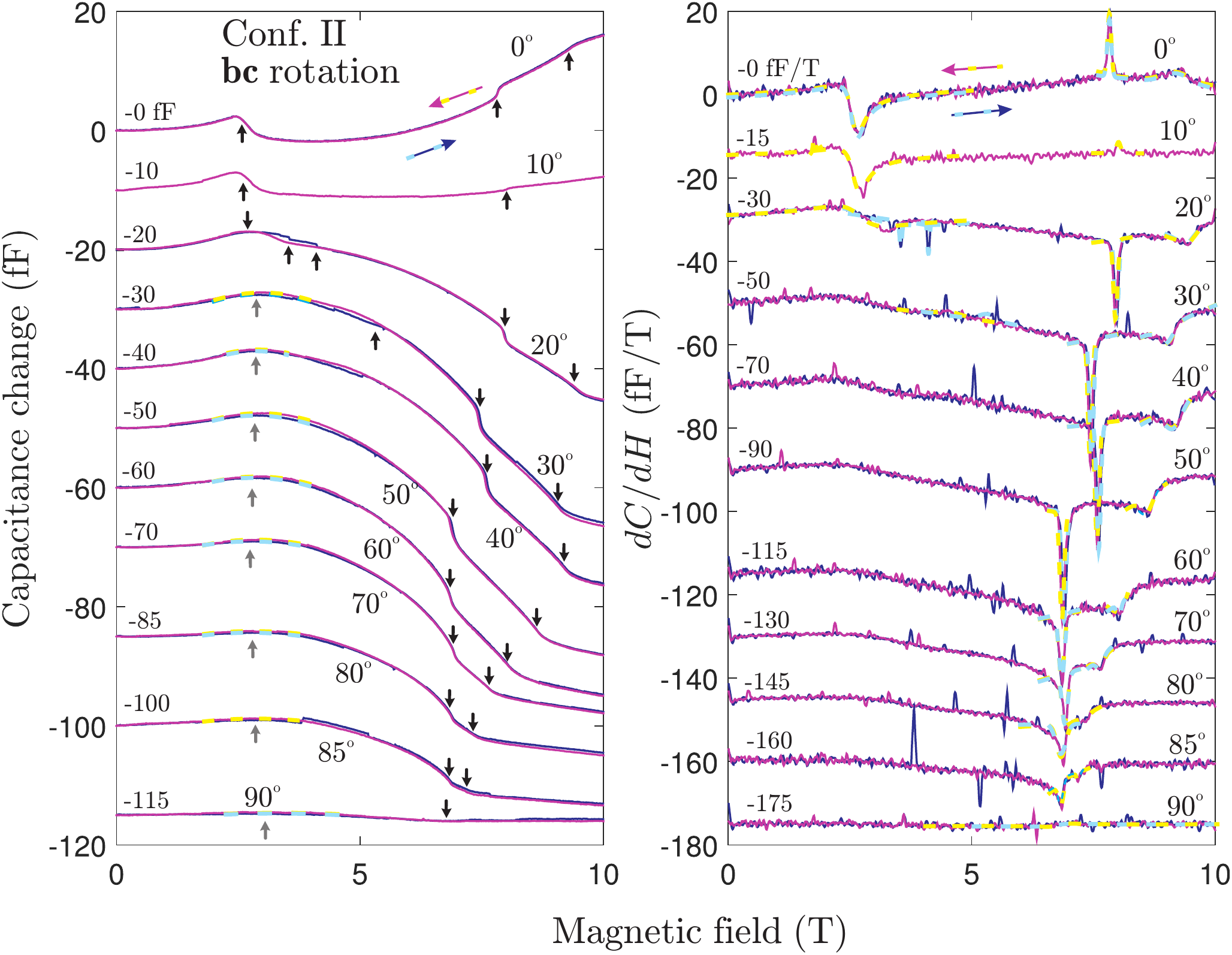}\\
  \caption{Evolution of the $\Delta C(H)$ torque signal as the magnetic field is
  rotated in the $\mathbf{bc}$ plane. The measurement geometry is configuration
  II.
  Selected torque curves and their derivatives at $T=0.2$~K are plotted as solid lines in the left and right panels correspondingly.
  The angles and offsets are indicated in the plots. Dashed lines indicate various anomalies approximated by
  Eqs.~(\ref{EQ:difflorenzfit}-\ref{EQ:parabola}) as described in the main text. Vertical arrows mark the obtained fields (black --- transitions,
  gray --- crossovers).}\label{FIG:BCcurves200}
\end{figure}

\begin{figure}
  \includegraphics[width=0.5\textwidth]{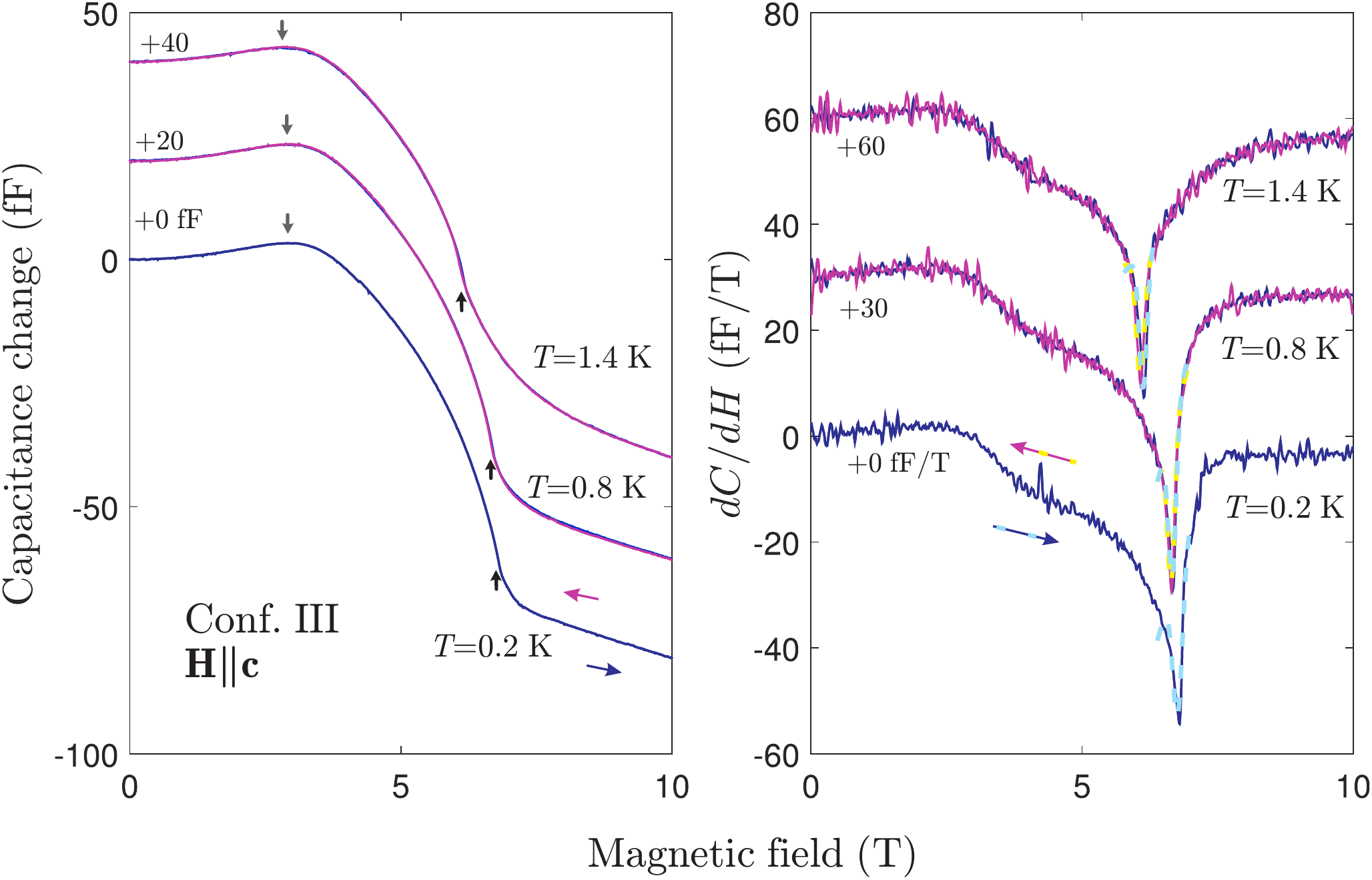}\\
  \caption{Selected $\Delta C(H)$ curves and their derivatives for $\mathbf{H}\parallel\mathbf{c}$ (configuration III).
  The temperatures and offsets are indicated in the plots.
  Dashed lines indicate various anomalies approximated by
  Eqs.~(\ref{EQ:gaussfit},\ref{EQ:parabola}) as described in the main text. Vertical arrows mark the obtained fields (black --- transitions,
  gray --- crossovers).}\label{FIG:Ccurves}
\end{figure}

\begin{figure}
  \includegraphics[width=0.5\textwidth]{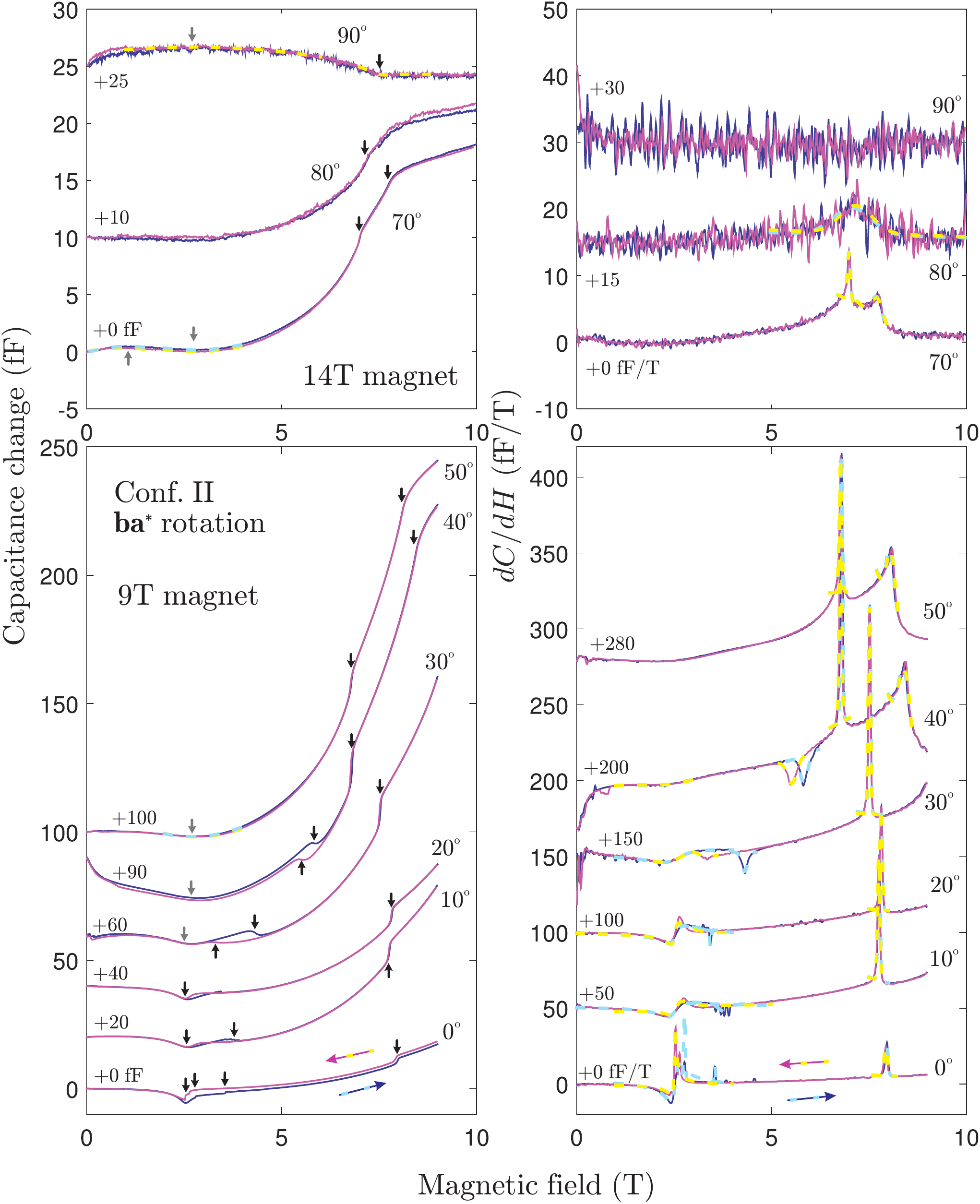}\\
  \caption{Evolution of the $\Delta C(H)$ torque signal as the magnetic field is
  rotated in the $\mathbf{ba}^{\ast}$ plane. The measurement geometry is configuration
  I.
  Selected torque curves and their derivatives at $T=0.2$~K are plotted as solid lines in the left and right panels correspondingly.
  The angles and offsets are indicated in the plots. Upper and lower set of panels contain the data recorded with 14~T or 9~T magnet,
  correspondingly. Dashed lines indicate various anomalies approximated by
  Eqs.~(\ref{EQ:difflorenzfit}-\ref{EQ:parabola}) as described in the main text. Vertical arrows mark the obtained fields (black --- transitions,
  gray --- crossovers).}\label{FIG:BAastcurves200}
\end{figure}

\begin{figure}
  \includegraphics[width=0.5\textwidth]{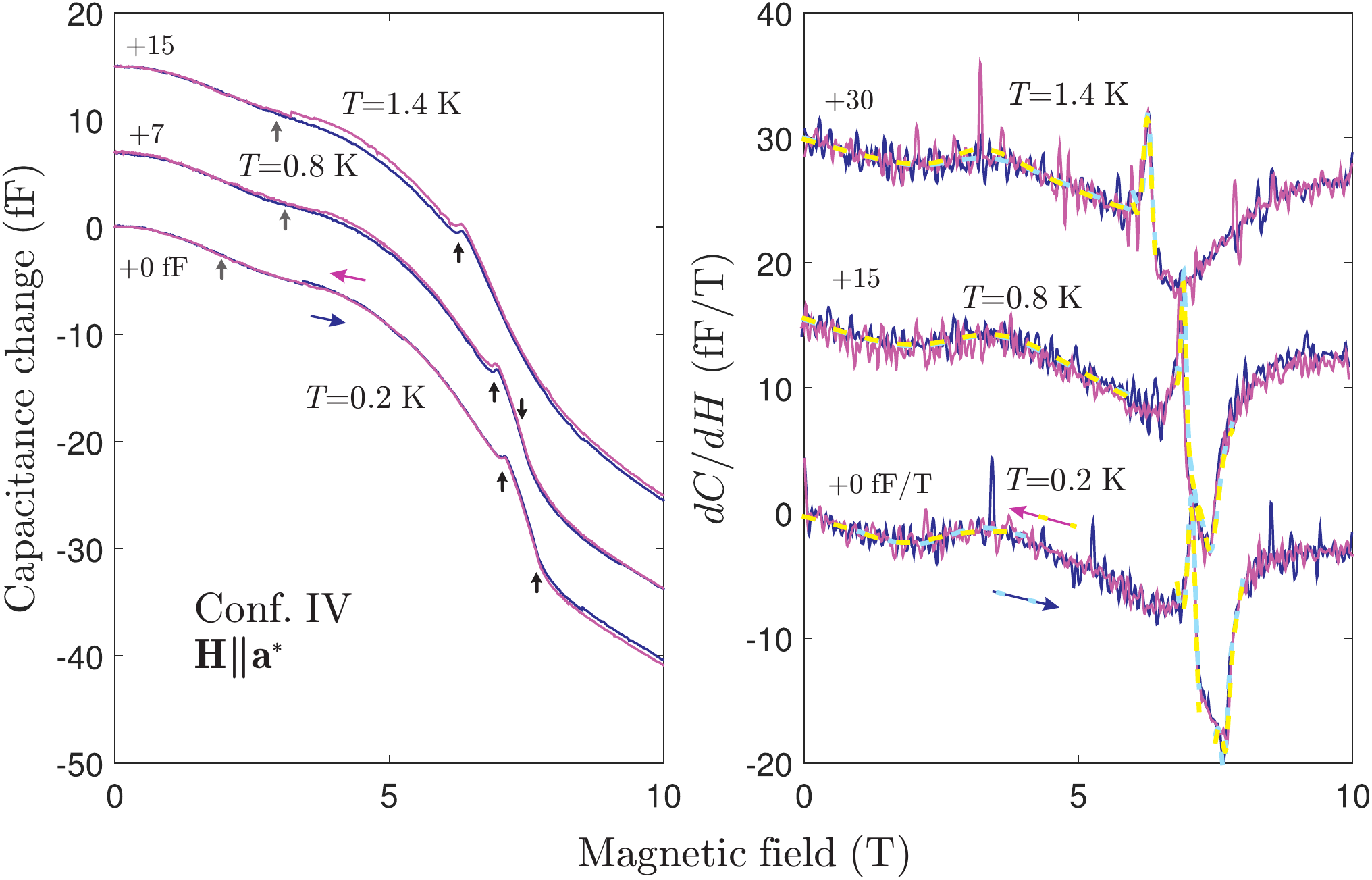}\\
  \caption{Selected $\Delta C(H)$ curves and their derivatives for $\mathbf{H}\parallel\mathbf{a}^{\ast}$ (configuration IV).
  The temperatures and offsets are indicated in the plots.
  Dashed lines indicate various anomalies approximated by
  Eqs.~(\ref{EQ:difflorenzfit},\ref{EQ:gaussfit}) as described in the main text. Vertical arrows mark the obtained fields (black --- transitions,
  gray --- crossovers).}\label{FIG:AAcurves}
\end{figure}

\begin{figure}
  \includegraphics[width=0.5\textwidth]{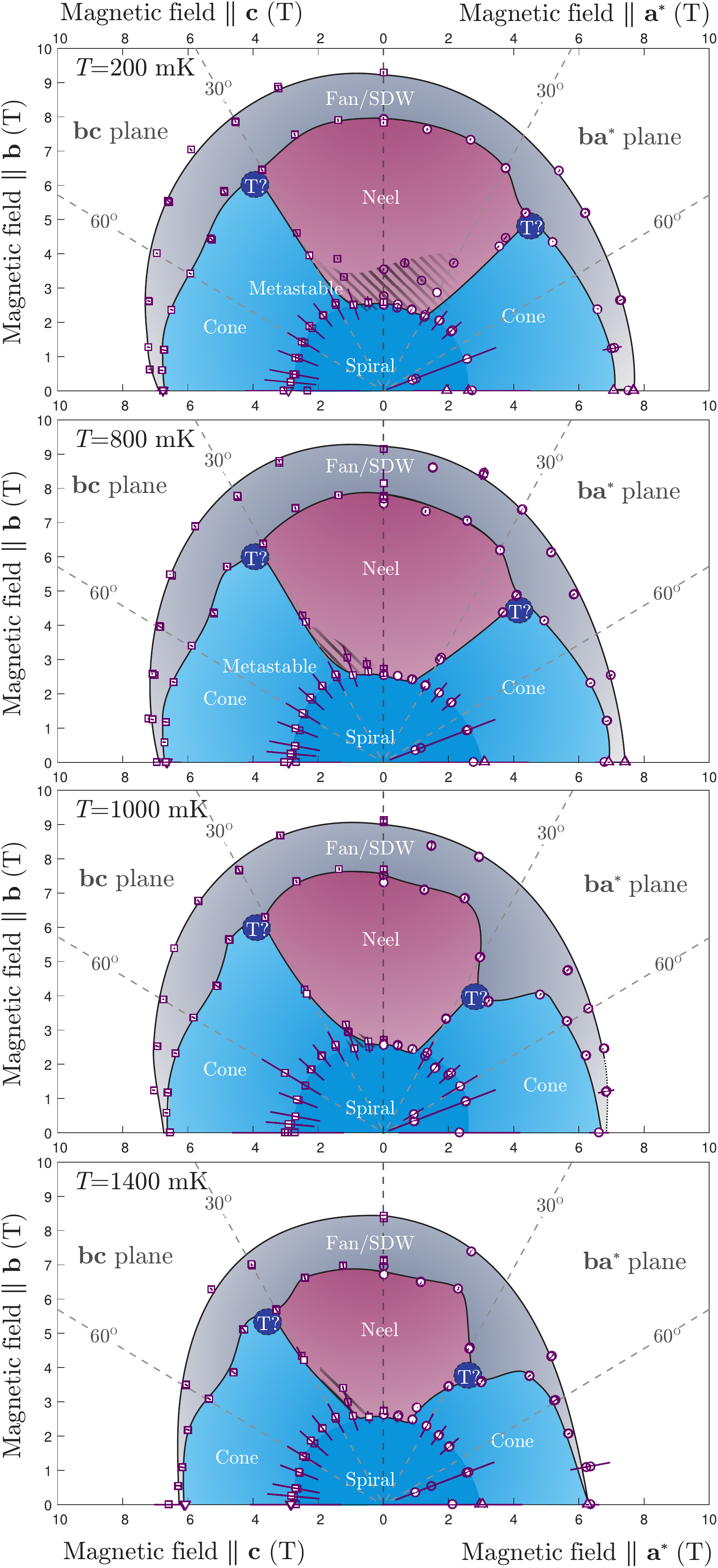}\\
  \caption{Angular phase diagram of linarite in $\mathbf{bc}$ and $\mathbf{ba}^{\ast}$ planes. Symbols correspond to the positions
  of the anomalies in the torque data; hollow symbols for sweeping the field up and filled symbols for sweeping the field down.
  Different symbol shapes correspond to different measurement configurations: circles, squares, downward and upward triangles ---
  for setups I, II, III and IV correspondingly. Lines are guide to the eye. Circles labeled with ``T?'' mark the regions where the tricritical point
  is expected.}\label{FIG:AngDiagAll}
\end{figure}

The first thing that is happening as the cantilever is rotated is
the shift of the anomalies positions. This is the manifestation of
shifting magnetic phase boundaries.
Second, the apparent amplitudes of the anomalies may change as well.
There are both intrinsic and extrinsic reasons for this. The
anomalies may indeed become less pronounced as certain phases become
suppressed and the corresponding order parameter vanishes. On the
other hand, the cantilever sensitivity depends on the geometry which
may or may not be favorable. For example, at $10^{\circ}$ tilt towards
$\mathbf{c}$ (see the corresponding curve in
Fig.~\ref{FIG:BCcurves200}) the forces acting on the cantilever
become rather compensated in the deflection direction, resulting in
a very weak signal. Consistently, the deflection of the cantilever
goes inwards or outwards for smaller or larger field tilts.
Nonetheless, in terms of transition-related anomalies the evolution
is smooth until $30^{\circ}$ where the sharp wiggle related to the
transition between spiral and commensurate states is gone.

At higher tilt angles this feature gives way to a broad maximum in
$\Delta C(H)$. This maximum continues to carry useful information on
the spin structure. The non-monotonous character of the curve
signals a competition between forces resulting from transverse and
longitudinal magnetization components in Eq.~(\ref{EQ:torquedef}).
As for a given run the geometry is fixed, the maximum (or minimum)
in the deflection signals the change of balance between
$\mathbf{M}\parallel \mathbf{H}$ and $\mathbf{M}\perp \mathbf{H}$
components, and hence a significant reorientation of the spin
structure. It looks much more like a crossover than a proper phase
transition, as the associated feature is quite broad. We can
empirically describe it as a simple parabola:

\begin{equation}\label{EQ:parabola}
    \Delta C(H)=b\pm\left(\frac{H-H_{0}}{\sigma}\right)^{2}.
\end{equation}

Again, $b$ is the purely empirical offset with no physical meaning,
while $H_{0}$ and $\sigma$ serve as the feature center and width
estimate. We can guess that such anomaly corresponds to a
transformation from a ``flat'' zero-field spin spiral into a cone phase
with significant polarization along the field direction. This
reorientation feature persists in the data all the way to the fully transverse field
geometry.

We can also resolve the anomalies corresponding to the boundaries of
the high field phase at least up to $85^{\circ}$. The enigmatic
``Fan/SDW'' phase persists, although it shrinks as the field comes
to the transverse orientation. In the fully transverse geometry with
$\mathbf{H}\parallel \mathbf{c}$ in configuration II the signal is
again dramatically reduced and it is impossible to draw any
conclusions about the presence of the high field phase. This
motivated us to use an additional geometry III, with
$\mathbf{H}\parallel \mathbf{c}$ being in the sensitive torquemeter
configuration. The results are shown in Fig.~\ref{FIG:Ccurves}. The
conclusion is that within the experimental resolution one observes
just one high field anomaly even at the lowest temperatures and the
high field phase is absent for $\mathbf{H}\parallel \mathbf{c}$
exact orientation.

A similar sequence of events is happening in case of magnetic field
tilt from $\mathbf{b}$ to $\mathbf{a}^{\ast}$ as shown in
Fig.~\ref{FIG:BAastcurves200}. The quantitative difference is that
the Ne\'{e}l phase is somewhat more robust in this case and holds
until $40^{\circ}$ tilt. As the low-tilt series of data were
measured in a machine with a 9~T magnet, we are also missing the
high field saturation anomaly in some of these curves, as it was
simply out of the accessible range. However, it finally appears
below 9~T as the tilt exceeds $30^{\circ}$. We are able to trace the
boundaries of the high field phase up to $70^{\circ}$; for higher
tilts the signal-to-noise screens the fine structure of the high
field anomaly. Again, this can be overcome by employing the
sensitive geometry IV with $\mathbf{H}\parallel\mathbf{a}^{\ast}$.
The results are shown in Fig.~\ref{FIG:AAcurves}. Surprisingly, in
this case we find the high field phase present and clearly
resolved.

The data from all the measurements at all the temperatures is
summarized in a series of angular phase diagrams present in
Fig.~\ref{FIG:AngDiagAll}. They will be discussed in detail in the
next section.

\section{Discussion}

The main result of this study is the angular phase diagrams in
Fig.~\ref{FIG:AngDiagAll}, which can be briefly summarized as
follows: the Ne\'{e}l phase is rather fragile and vanishes at
approximately $40^{\circ}$ tilt from the $\mathbf{b}-$axis, while
the enigmatic ``Fan/SDW'' phase that precedes full saturation turns
out to be robust and may indeed persist even in the transverse
magnetic field orientation. Both findings are qualitatively
consistent with the theoretical predictions of Cemal~\emph{et
al.}~\cite{CemalEnderle_PRL_2018_Linarite}. In particular, in the
close to $a$ direction of magnetic field we do observe a
non-vanishing high field phase, in agreement with the direct
observations of the ``Fan'' state by Cemal~\emph{et al.}.
Unfortunately, the static uniform magnetization measurements do not
provide us with any microscopic information, and thus it is not
possible to differentiate between the ``Fan'' and ``SDW''
possibilities from our set of data to extend this comparison
further.

Fig.~\ref{FIG:AngDiagAll} also plots the crossover from ``flat''
zero-field spiral to the partially polarized cone state. As
discussed above, on this line the structure becomes predominantly
polarized along the field around $2.8-3$~T, in agreement with
neutron diffraction data~\cite{CemalEnderle_PRL_2018_Linarite}.
While in the neutron diffraction data this microscopic change of
structure is rather sharp and pronounced, in torque magnetometry
measurements it appears as a broad crossover. This loosely defined
crossover field is replaced by a sharp transition in the narrow
angular range supporting the collinear Ne\'{e}l phase. Metastability
effects stress the first order nature of that transition.
Interestingly, in the exact $\mathbf{H}\parallel\mathbf{b}$
orientation history-dependent behavior is confined to the lowest
temperatures, while with the deflection towards the
$\mathbf{c}$-axis they start to proliferate and become present in
the whole temperature range of the study. As soon as the Ne\'{e}l
phase ceases to exist, any history dependent behavior disappears.

An important observation is that the field at which the flat spin
spiral structure is transformed, either through a crossover or a
phase transition, is nearly the same for all orientations. This
tells us that the same energy scale is at play, that is the main
easy axis anisotropy ($\vec{\xi}_{1}\parallel \mathbf{b}$ direction
in Fig.~\ref{FIG:structure}). On the other hand, the Ne\'{e}l phase
is supposedly stabilized by the smaller anisotropy constant
(associated with the $\vec{\xi}_2$
direction)~\cite{CemalEnderle_PRL_2018_Linarite}. Thus, knowing the
critical angles at which the collinear phase disappears may be
essential to get an estimate of both anisotropy energies.

An interesting minor detail is the behavior of ``triple'' points
separating the Ne\'{e}l, high field and cone phases. Although we do
not have enough angular resolution to locate these points precisely
(their possible locations are indicated by large circles in
Fig.~~\ref{FIG:AngDiagAll}), it seems that around these points the
stability of the high field phase is enhanced. This behavior is
particularly pronounced at higher temperatures.

Another minor point concerns the intermediate small pocket of
``phase~III''~\cite{Willenberg_PRL_2016_LinariteSDWs,PovarovFeng_PRB_2016_LinariteMF}
which is found at higher temperatures for $\mathbf{H}\parallel
\mathbf{b}$ (as in Fig.~\ref{FIG:Bdiag}). In our experiments it
could not be clearly resolved in any other orientations and is
therefore not indicated in the Fig.~\ref{FIG:AngDiagAll} phase
diagrams.

\section{Conclusions}

The complex orientational magnetic phase diagram of linarite
reflects a subtle competition between anisotropy terms in the
magnetic Hamiltonian. Nevertheless, it is not at all inconsistent
with the ``big picture'' of competing quantum phases in the
simplified $J_1-J_2$ Heisenberg model. On a qualitative level, our
findings are consistent with the mean field model of
Ref.~\cite{CemalEnderle_PRL_2018_Linarite}. Further theoretical work
is needed to enable a quantitative comparison.

\acknowledgments

This work was supported by Swiss National Science Foundation,
Division II.

\bibliography{d:/The_Library}
\end{document}